\documentclass[submitting]{nst}

\usepackage{subfigure,dcolumn}
\usepackage{epstopdf}
\usepackage{mhchem}
\usepackage{upgreek}
\usepackage{multirow}
\usepackage{booktabs}
\usepackage{graphicx}

\begin{document}

\title{Non-extensive statistical distributions of charmed meson production in Pb-Pb and pp($\overline{\text{p}}$) collisions}
\thanks{Supported by National Key Research and Development Program of China with Grant No. 2018YFE0205200 and 2018YFE0104700, National Natural Science Foundation of China with Grant No. 11890712 and 12061141008, the Strategic Priority Research Program of CAS with Grant No. XDB34030000 and Anhui Provincial Natural Science Foundation with Grant No. 1808085J02.}

\author{Yuan Su}
\affiliation{State Key Laboratory of Particle Detection and Electronics, University of Science and
Technology of China, Hefei 230026, China}
\affiliation{Department of Modern Physics, University of Science and Technology of China, Hefei 230026, China }

\author{Xiaolong Chen}
\affiliation{State Key Laboratory of Particle Detection and Electronics, University of Science and
Technology of China, Hefei 230026, China}
\affiliation{Department of Modern Physics, University of Science and Technology of China, Hefei 230026, China }

\author{Yongjie Sun}
\affiliation{State Key Laboratory of Particle Detection and Electronics, University of Science and
Technology of China, Hefei 230026, China}
\affiliation{Department of Modern Physics, University of Science and Technology of China, Hefei 230026, China }

\author{Yifei Zhang}
\affiliation{State Key Laboratory of Particle Detection and Electronics, University of Science and
Technology of China, Hefei 230026, China}
\affiliation{Department of Modern Physics, University of Science and Technology of China, Hefei 230026, China }

%\email[Corresponding author, ]{machunwang@126.com}

\begin{abstract}
The mid-rapidity transverse-momentum spectra of charmed meson in Pb-Pb and pp($\overline{\text{p}}$) collisions are analyzed by Tsallis-Pareto distribution derived from non-extensive statistics. We perform a uniform description for both small and large systems over a wide range of collision energy and hadron transverse momentum. By establishing the relation between the event multiplicity and the Tsallis parameters, we observe that there is a significant linearity between thermal temperature and Tsallis-$q$ parameter in Pb-Pb collisions at $\sqrt{s_{NN}}$ = 2.76 TeV and 5.02 TeV. And the slope of $T$ - ($q-1$) parameter is positively correlated with the hadron mass. In addition, the charmed mesons have higher thermal temperature than the light hadrons at the same $q-1$, indicating that charmed flavor requires a higher temperature to achieve the same degree of non-extensive as light flavors in the heavy-ion collision. A same fit is performed to the transverse momentum spectra of charmed meson in pp($\overline{\text{p}}$) collision over a large energy range with Tsallis-Pareto distribution. It is found that the thermal temperature increases with system energy while the $q$ parameter shows a saturated trend and stops at pp($\overline{\text{p}}$) limit, $q-1$ = 0.142 $\pm$ 0.010. Meanwhile, the results of most peripheral Pb-Pb collisions are found to approach the pp($\overline{\text{p}}$) limit, which suggests that less medium effect and more in similarity with pp($\overline{\text{p}}$) collisions are found in more peripheral heavy-ion collisions.
\end{abstract}


\maketitle

\section{Introduction}\label{sec.I}

According to the Big Bang theory, under the condition of extremely high temperature and high energy density at the early stage of the universe, Quark-Gluon Plasma (QGP), a new form of matter, could be generated by the release of quarks and gluons that were bound in hadrons due to strong interaction. The masses of the heavy flavor quarks, $m_c\thicksim$ 1.3 GeV/$c^2$, $m_b\thicksim$ 4.8\ GeV/$c^2$, are larger than that of light quarks and quantum chromodynamics (QCD) energy scale ($\varLambda _{QCD}$). Therefore, the generation of heavy flavors requires large enough energy and momentum transfer by initial hard scatterings in heavy-ion collisions, and can be calculated by perturbation quantum chromodynamics~\cite{b1,b2}. Heavy flavor quarks have a high probability of experiencing a relatively complete evolution process of QCD materials, so heavy flavor quarks are ideal probes to study QGP properties in heavy-ion collisions. In particular, it is crucial to study the interaction between heavy flavor quarks and medium by measuring the nuclear modification factor, flow and production yield of charmed meson~\cite{b3,b4,b5,b6,b7,b8,b9,b10,b11,b12}.

The transverse momentum ($p_T$ ) spectra of identified particles provide valuable data for discoveries based on relativistic heavy-ion collisions~\cite{b13,b14}. Boltzmann Gibbs Blast-Wave model with a compact set of parameters of temperature ($T$) and flow velocity ($\beta$) had proved that the spectral shape is sensitive to the dynamics of the nucleus-nucleus (AA) collisions ~\cite{b13,b15} and can be used to describe the transverse momentum distributions of light flavor hadrons with different masses. Moreover, the Tsallis blast-wave model, including  fluctuation of initial condition for the hydrodynamic evolution on an event-by-event basis, was used to study the $\pi^{\pm}$, $K^{\pm}$, $p$($\overline{p}$), $\phi$, $\Lambda(\overline{\Lambda})$ and $\Xi^{-}(\overline{\Xi}^{+})$ spectra in Au-Au collisions at 200 GeV~\cite{b16}. They found that the average transverse flow velocity $\langle\beta\rangle$ and the temperature at the kinetic freeze-out $T_{fro}$ increased with centrality while the non-extensive parameter $q$ showed the opposite behaviour. Recently, the thermodynamical formulation of non-extensive statistics was applied to the data of all light hadrons, and a strong grouping phenomenon in the $T$ - ($q-1$) parameter space was considered to be a probe for the study of QGP in small and large systems. On the one hand, the non-extensive temperature converges to $T\approx$ 0.144 $\pm$ 0.010 GeV~\cite{b17}, which is consistent with the QCD results~\cite{b18,b19}, indicating that the collisional system may have undergone QGP evolution. On the other hand, the non-extensive parameter $q \not=$ 1, indicating that the conclusion is entirely derived from the non-extensive properties. Although the implications and understanding of the consequences of such an application to particle production are still under investigation, the function is relatively easy to understand~\cite{b20,b21,b22}. Heavy flavors are expected to behave differently from light flavors, which may result in different non-extensive thermal parameters. The same idea with non-extensive statistical distribution function is extended to fit the charmed meson spectra recently measured in ALICE and STAR experiments.

In this article, we present the Tsallis-Pareto distribution formula and explain the physical origin of the relationship between the parameters $T$ and $q$ in section~\ref{sec.II}. The analysis procedure and fit results of charmed meson and the thermal temperature after transverse flow correction are also given in this section. In section~\ref{sec.III}, the results for 0-10\% (central) and 30-50\% (semi-peripheral) centrality bins in Pb-Pb collisions at 2.76 TeV and for 0-10\%, 30-50\% and 60-80\% (peripheral) centrality bins in Pb-Pb collisions at 5.02 TeV are presented, respectively. In the same section, the results obtained in Pb-Pb collisions are compared with the values, $T$ and $q$, measured in proton-(anti)proton (pp($\overline{\text{p}}$)) collisions over a large energy span. The results and conclusions are summarized in section~\ref{sec.IV}.

\section{Tsallis-Pareto distribution and its fit to charmed meson spectra}\label{sec.II}
In high-energy particle collisions, much work is devoted to the study of the transverse momentum distributions of outgoing particles. At low-$p_T$ regime of the spectra, the traditional exponential distributions can be used to describe the spectral shape and the formula, assuming vanishing chemical potential at high energies, is given as
\begin{equation}
\label{eq1}
f\left( p_T \right) \thicksim \exp \left( -\frac{m_T}{T} \right) ,
\end{equation}
where $m_T$ = $\sqrt{p_T^2+m^2}$ is the transverse mass and m is the rest mass of the given hadron species, $T$ is the corresponding temperature. Instead of Eq.(\ref{eq1}), a power law distribution has been used previously in high-energy physics to better characterize the spectra with $p_T\gtrsim$ 3 GeV/c. Tsallis-Pareto distribution derived from non-extensive statistics, which is a generalization of the traditional Boltzmann-Gibbs theory, is supposed to describe the full $p_T$ spectrum of hadrons. So far the transverse momentum dependence of identified hadrons production from recently published data was successfully described by non-extensive statistics~\cite{b23}. And as a theoretical limit case $q\to1$, the exponential and logaritm function can be obtained~\cite{b24,b25,b26}. This new stream in the foundation of statistical mechanics was initiated by Tsallis' proposal of a non-extensive entropy in 1988~\cite{b24}. Non-extensive entropy (Tsallis entropy) is a generalization of the traditional Boltzmann-Gibbs entropy, and it formed the basis of non-extensive statistics.

The same Tsallis-Pareto distribution within the non-extensive statistics as in~\cite{b17} is used in the current analyses:
\begin{equation}
\label{eq2}
\begin{aligned}
\frac{dN}{2\pi p_Tdp_T}=Am_T\left[ 1+\frac{q-1}{T_q}\left( m_T-m \right) \right] ^{-\frac{q}{q-1}},
\end{aligned}
\end{equation}
where $A$ is a normalizing factor and can reflect the production yield of the hadron pT spectrum. $T_q$ is the temperature in the non-extensive statistical theory and the subscript $q$ will be omitted for brevity in the following. Note that $T_q$ can be different from temperature $T$ in Eq.(\ref{eq1}), but its physical meaning should be the same in a limiting case, $q\to1$. 

The correlations between $T$ and $q$ parameters have been revealed in some earlier studies~\cite{b27,b28,b29,b30}. Furthermore, the charged particle multiplicity can be derived from Tsallis-Pareto distributed transverse momentum, and concrete application to experimental data results in a negative binomial distribution parameter $k\ \thicksim\ \mathcal{O}\left( 10 \right) $~\cite{b31,b32,b33,b34,b35,b36,b37,b38}. In addition, many studies revealed that the measurement of event-by-event multiplicity and its fluctuation is of great significance. They found that the yield of strange hadron is positively related to multiplicity and a long-range correlation can be observed in a small collisional systems as the multiplicity increases~\cite{b39,b40,b41,b42,b43}. This study is specific to measure the parameters relationship of heavy flavor hadrons in the $T$ - ($q-1$) parameter space using the non-extensive statistics described above. For simplicity, the fluctuations in the number of the produced particles can be explained in a one dimensional relativistic gas model~\cite{b44}, and the Tsallis parameters under consideration given as:
\begin{equation}
\label{eq3}
T=\frac{E}{\langle M\rangle},
\end{equation}
\vspace{-6mm}
\begin{equation}         
\label{eq4}
q=1-\frac{1}{\langle M\rangle}+\frac{\varDelta M^2}{\langle M\rangle^2},
\end{equation}
Here $M$ is the number of particles at an associated energy E.

In the thermodynamic picture, the relation between $T$ and $q$ can be obtained from Eq.(\ref{eq3}) and Eq.(\ref{eq4}), assuming that the relative size of multiplicity fluctuations are constant as in~\cite{b17}:
\begin{equation}
\label{eq5}
\frac{\varDelta M^2}{\langle M\rangle^2}\equiv \zeta ^2,
\end{equation}
so that we have the following formula  in the current analyses
\begin{equation}
\label{eq6}
T=E\left( \zeta ^2-\left( q-1 \right) \right).
\end{equation}
which is used to measure the relationship between the Tsallis parameters and event multiplicity in charmed meson production for both small and large systems over a wide range of collision energy and hadron transverse momentum, and compare with corresponding results of light hadrons.

\subsection{Implement Tsallis-Pareto distribution into charmed meson spectra}\label{sec.3.1}
\begin{figure}[!htb]
\includegraphics
[width=0.9\hsize]{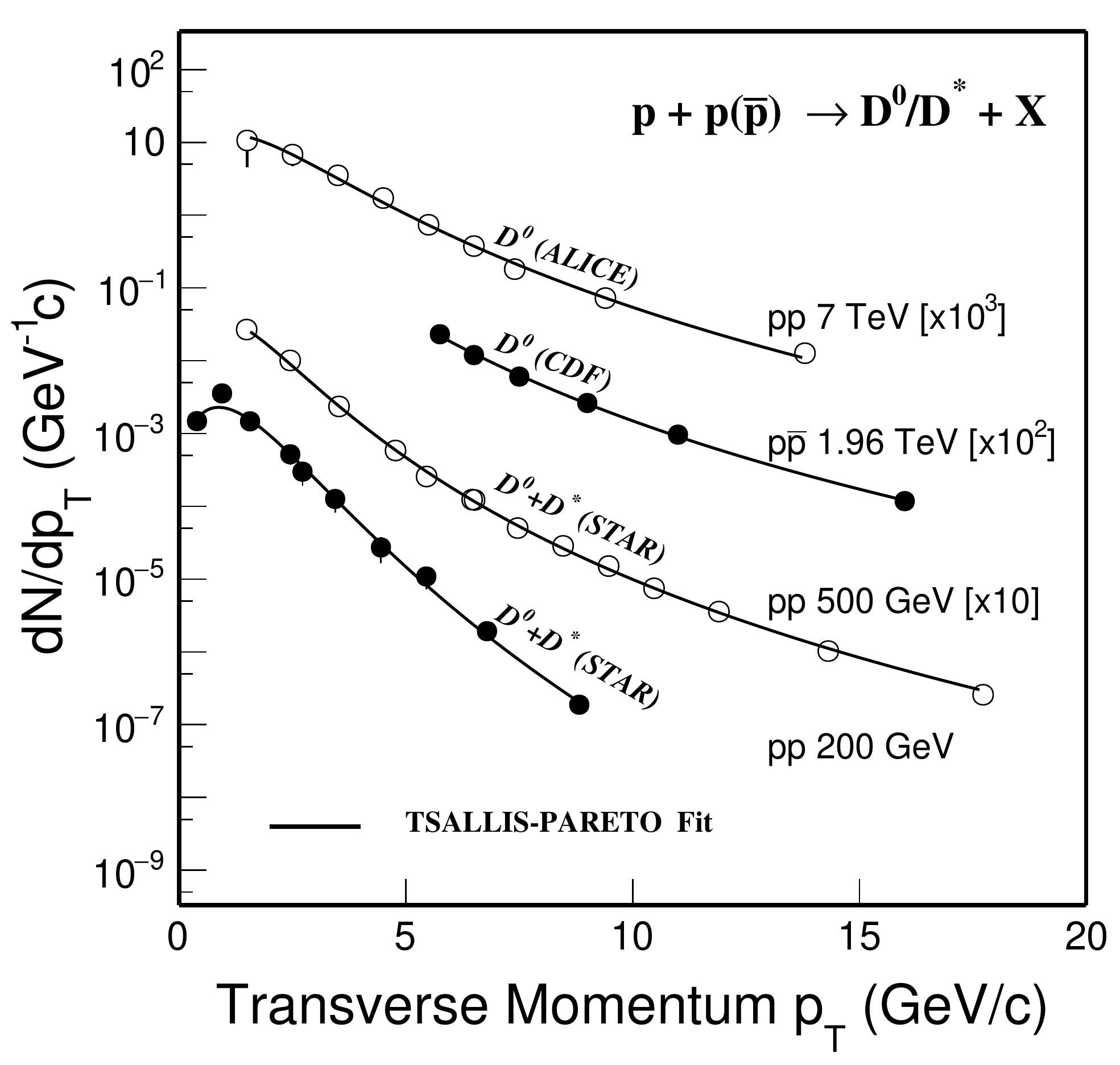}
\caption{Transverse momentum distributions $dN/dp_{T}$ of $D^0$/$D^{*}$ in pp($\overline{\text{p}}$) collisions at 200 GeV, 500 GeV, 1.96 TeV and 7 TeV from bottom to top. The solid curves are results from the Tsallis-Pareto fit. Error bars are quadratical sum of statistical and systematic uncertainties, and data are scaled by $10^{n}$ factors for better visibility.}
\label{fig1}
\end{figure}

In the present study we analyze the transverse momentum dependence of charmed meson production in pp($\overline{\text{p}}$) and AA collisions measured by the ALICE, CDF and STAR collaborations~\cite{b45,b46,b47,b48,b49,b50,b51,b52}. We find that the charmed mesons data from small and large collision system with a broad selection criteria can be investigated simultaneously. $D^0$ and $D^{*}$ spectra are measured by STAR Collaboration in pp collision at 200 GeV for 0.4 $< p_T <$ 8.8 GeV/c and at 500 GeV for 1.5 $< p_T <$ 17.7 GeV/c , respectively. $D^0$ spectra are measured by CDF Collaboration in p$\overline{\text{p}}$ collision at 1.96 TeV for 5.8 $< p_T <$ 16.0 GeV/c and by ALICE Collaboration in pp collision at 7 TeV for 1.5 $< p_T <$ 13.8 GeV/c, respectively. Figure \ref{fig1} shows above $p_T$ spectra of D mesons together with our fit results in four selected energy (200 GeV, 500 GeV, 1.96 TeV and 7 TeV) in pp($\overline{\text{p}}$) collisions. The fit parameters and $\chi^2/ndf$ are tabulated in TABLE~\ref{table1}. The solid curves from Tsallis-Pareto distributions describe the data well. Error bars denote quadratical sum of statistical and systematic uncertainties. Data are scaled by $10^{n}$ factors for better visibility.

\begin{figure}[!htb]
\includegraphics
[width=0.9\hsize]{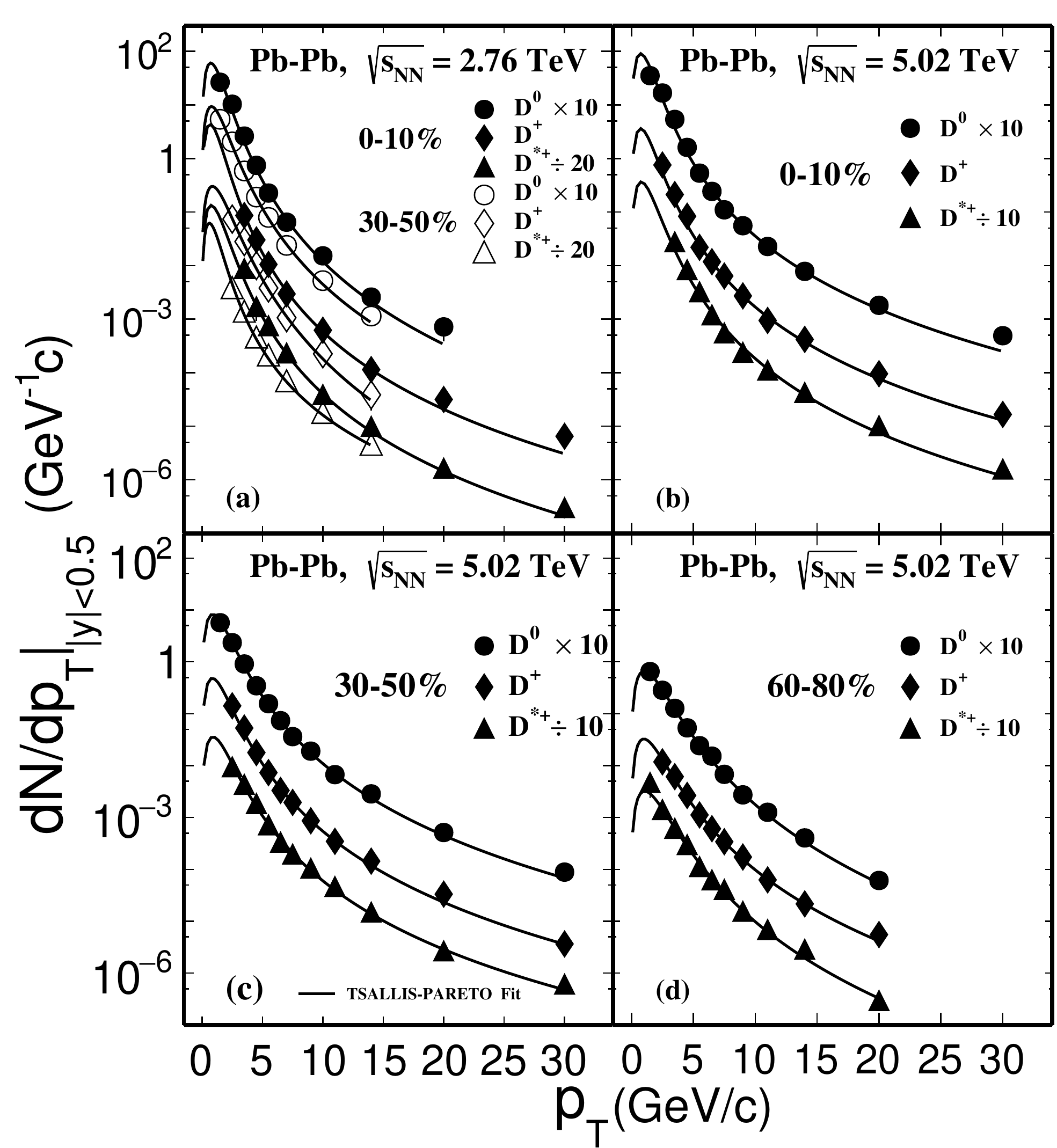}
\caption{Transverse momentum distributions $dN/dp_{T}$ of $D^0$, $D^+$ and $D^{*+}$ for different centrality bins in Pb-Pb collisions at 2.76 TeV (a) and 5.02 TeV (b) , (c) , (d) , where the production yields are scaled by factors for visibility. The vertical bars represent quadratical sum of statistical and systematic errors; symbols are placed at the centre of the bin. The detailed descriptions are shown in section~\ref{sec.3.1}.}
\label{fig2}
\end{figure}

In Fig.~\ref{fig2}, the fits are applied to the transverse momentum distributions of charmed mesons for different centrality bins in AA collisions~\cite{b49,b50,b51,b52}. The fits of the transverse momentum distributions of prompt $D^0$, $D^+$ and $D^{*+}$ mesons in Pb-Pb collisions at 2.76 TeV are shown in Fig.~\ref{fig2}(a). Solid circles, diamonds and  triangles represent $D^0$, $D^+$ and $D^{*+}$, respectively. The 0-10\% (solid) and 30-50\% (hollow) centrality bins are drawn on the same panel (a)~\cite{b49}, where the $D^{0}$ and $D^{*+}$ production yields at 0-10\% (30-50\%) are scaled by the factors 10 and 0.05, respectively. For 0-10\% centrality bin, fits are performed in the range of 0 $<$ $p_T$ $<$ 20.0 GeV/c for $D^0$ and in the range of 0 $< p_T <$ 30.0 GeV/c for $D^+$ and $D^{*+}$, respectively. For 30-50\% centrality bin, the same fitting procedures are performed in the range of 0 $< p_T <$ 14.0 GeV/c for $D^0$, $D^+$ and $D^{*+}$. The vertical bars represent the sum of statistical uncertainties and systematic uncertainties. TABLE~\ref{table1} shows that the fit parameters of charmed meson spectra at 0-10\% centrality bin have smaller uncertainties with more data points than those at 30-50\% centrality bin at 2.76 TeV.

% \usepackage{booktabs}
% \usepackage{multirow}
\begin{table*}[]\footnotesize

\setlength{\tabcolsep}{-6mm}
\centering \caption{Values of parameters from Tsallis-Pareto fit to charmed mesons spectra in Pb-Pb (pp($\overline{\text{p}}$)) collisions. The uncertainties are from the fit.}\label{runResult}
\centering

\setlength{\tabcolsep}{5.5mm}{
\begin{tabular}{@{}ccccccc@{}}
\toprule
$\sqrt{s_{NN}}$ (GeV) & Centrality & Charmed meson & A          & T         & q        & $\chi ^2/ndf$       \\ \midrule
Pb-Pb, 2760  & 0-10\%    & $D^0$     & $1.279\pm 0.516$ & $0.239\pm0.030$ & $1.166\pm 0.014$ & $5.15/6$ \\
          &            & $D^+$     & $1.052\pm 0.350$ & $0.201\pm 0.024$ & $1.180\pm 0.013$ & $9.13/7$ \\
          &            & $D^{*+}$     & $0.515\pm 0.212$ & $0.240\pm0.031$ & $1.179\pm 0.012$ & $2.9/7$ \\
          & 30-50\%    & $D^0$     & $0.178\pm 0.076$ & $0.278\pm0.041$ & $1.169\pm0.020$ & $0.88/5$ \\
          &            & $D^+$     & $0.052\pm 0.025$ & $0.322\pm 0.055$ & $1.151\pm0.025$ & $0.51/5$ \\
          &            & $D^{*+}$     & $0.084\pm 0.045$ & $0.250\pm0.049$ & $1.205\pm 0.025$ & $0.34/5$ \\
Pb-Pb, 5020  & 0-10\%    & $D^0$     & $1.957\pm 0.551$ & $0.240\pm 0.020$ & $1.187\pm0.008$ & $14.95/9$ \\
          &            & $D^+$     & $0.768\pm0.277$ & $0.245\pm 0.025$ & $1.190\pm 0.008$ & $5.16/9$ \\
          &            & $D^{*+}$     & $0.675\pm 0.209$ & $0.258\pm 0.026$ & $1.184\pm0.010$ & $6.12/9$ \\
          & 30-50\%    & $D^0$     & $0.145\pm 0.040$ & $0.328\pm 0.026$ & $1.175\pm 0.008$ & $2.12/9$ \\
          &            & $D^+$     & $0.088\pm 0.024$ & $0.311\pm 0.024$ & $1.179\pm 0.007$ & $5.35/9$ \\
          &            & $D^{*+}$     & $0.057\pm 0.020$ & $0.331\pm0.034$ & $1.185\pm 0.008$ & $3.76/9$ \\
          & 60-80\%    & $D^0$     & $0.010\pm 0.003$ & $0.427\pm 0.042$ & $1.151\pm 0.012$ & $2.52/8$ \\
          &            & $D^+$     & $0.005\pm 0.001$ & $0.402\pm 0.043$ & $1.170\pm 0.013$ & $2.21/8$ \\
          &            & $D^{*+}$     & $0.004\pm 0.002$ & $0.430\pm 0.069$ & $1.156\pm 0.017$ & $4.48/8$ \\
pp, 200       &         & $D^0+D^{*}$     & 3.8e-4 $\pm$ 5.3e-5 & $0.322\pm 0.022$ & $1.081\pm 0.011$ & $4.39/7$ \\
pp, 500      &          & $D^0+D^{*}$     & 6.8e-4 $\pm$ 1.7e-4 & $0.310\pm 0.020$ & $1.132\pm 0.006$ & $4.13/10$ \\
p($\overline{\text{p}}$), 1960     &            & $D^0$     & 1.8e-3$\pm$1.7e-3 & $0.386\pm 0.058$ & $1.143\pm 0.011$ & $0.87/3$ \\
pp, 7000       &        & $D^0$     & 1.6e-3 $\pm$ 7.6e-4 & $0.494\pm 0.062$ & $1.139\pm 0.023$ & $1.75/6$ \\ \bottomrule
\end{tabular}
}
\label{table1}
\end{table*}

The transverse momentum distributions $dN/dp_{T}$ of $D^0$ (solid circles), $D^+$ (diamonds) and $D^{*+}$ (triangles) mesons in 0-10\%, 30-50\% and 60-80\% centrality bins in Pb-Pb collisions at 5.02 TeV are shown in Fig.~\ref{fig2}(b), Fig.~\ref{fig2}(c) and Fig.~\ref{fig2}(d), respectively~\cite{b50}. The vertical bars represent quadratical sum of statistical and systematic uncertainties, symbols are placed at the centre of the bin. The solid curves from Tsallis-Pareto distributions describe the data well. For visibility, the $D^0$ and $D^{*+}$ at three different centrality bins are scaled by a factor of 10 and $1/10$, respectively. To more physically constrain the $D^{+}$ and $D^{*+}$ yield at $p_{T}$ = 1.5 GeV/c, we apply a $D^{+}$/$D^0$ and $D^{*+}$/$D^0$ ratio around 0.5 from PYTHIA and perform the fit. The ratio obtained from PYTHIA is consistent with the experimental data~\cite{b50}. The fit $T$, $q$, $A$ parameters and  $\chi^2/ndf$ values are tabulated in TABLE~\ref{table1}. The $T$, $q$ parameters after transverse flow corrections are shown in Fig.~\ref{fig5}. In addition, we also perform the same Tsallis-Pareto fits to the transverse momentum spectra of $\pi^{\pm}$, $K^{\pm}$ and $p$($\overline{p}$) at 0-5\%, 5-10\%, 10-20\%, 20-30\%, 30-40\%, 40-50\%, 50-60\%, 60-70\%, 70-80\%, and 80-90\% centrality bins in Pb-Pb collisions at 2.76 TeV and 5.02 TeV~\cite{b51,b52}. The fits parameters after transverse flow corrections are also shown in Fig.~\ref{fig5}. The $\phi$, $\Lambda^0$, $\Xi$ in different centralities at 2.76 TeV are studied using the same method. Point-by-point statistical and systematic uncertainties are added as a quadratic sum when we perform these fits. 

\subsection{Thermal temperature with flow correction}\label{sec.3.2}
\begin{figure}[!htb]
\includegraphics
  [width=0.9\hsize]{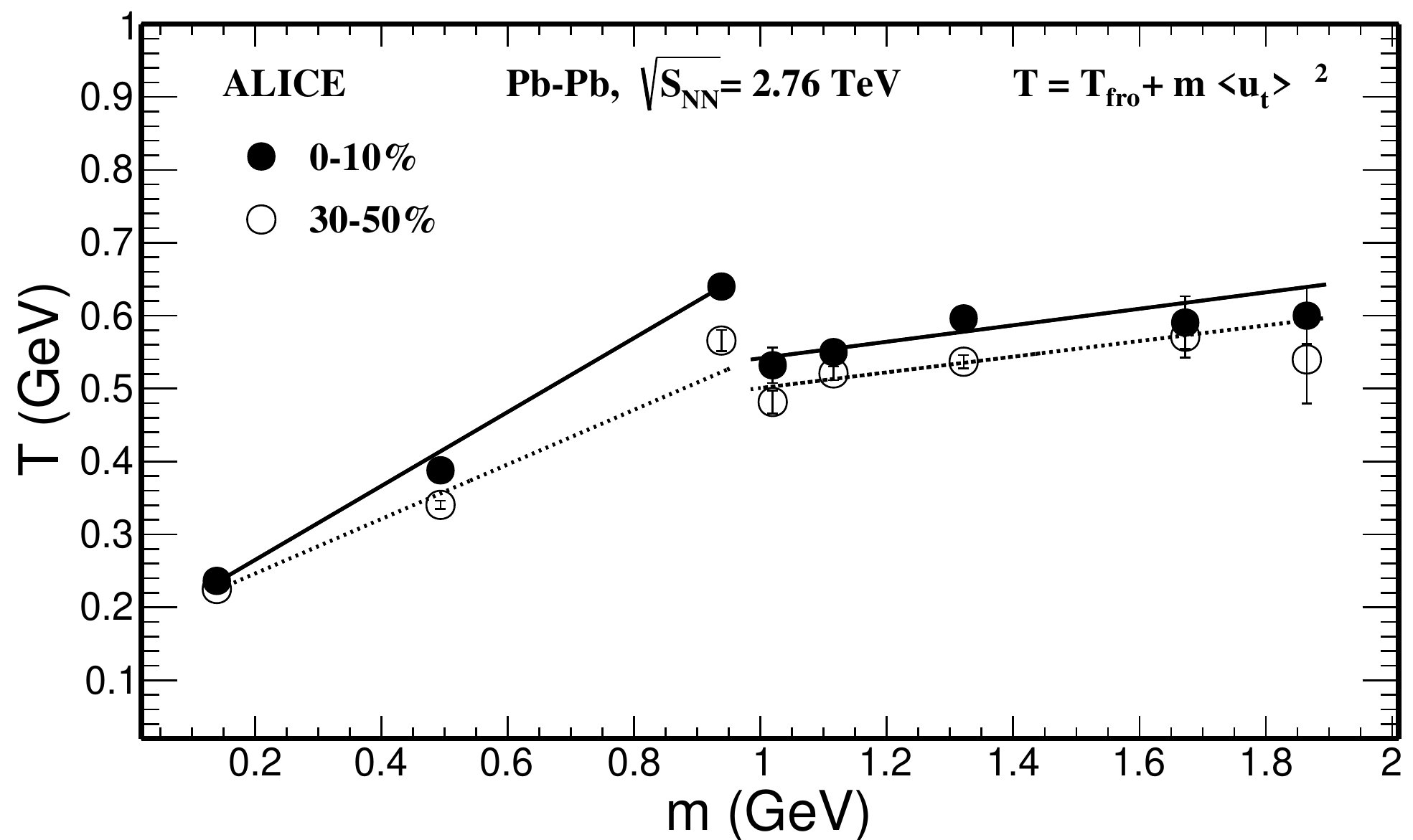}
\caption{The effective temperature as a function of hadron mass for 0-10\% (solid circles) and 30-50\% (hollow circles) in Pb-Pb collisions at 2.76 TeV. The lines are fits from Eq.(\ref{eq7}).}
\label{fig3}
\end{figure}

The phenomenological model can describe almost all hadronic spectra by starting out with thermalization and collective flow as the basic assumption~\cite{b15}. The mass dependence of the effective temperature $T$ has been described by introducing a Gaussian parameterization~\cite{b53,b54,b55,b56} and can be interpreted as the existence of a radial flow. The radial flow velocity, which generated by violent nucleon-nucleon collisions in two colliding nuclei and developed both in the QGP phase and in hadronic rescatterings, increases the transverse momentum of particles proportional to their mass~\cite{b15,b57,b58}. Many models attempt to investigate the radial flow~\cite{b59}, one more radial flow picture~\cite{b15,b58} we can use in the present analysis is
\begin{equation}
\label{eq7}
T=T_{fro}+m \langle u_t\rangle^2 ,
\end{equation}
where $T_{fro}$ is a hadron kinetic freeze-out temperature and $\langle u_t\rangle$ is a measure of the strength of the (average radial) transverse flow. The connection  between the averaged transverse velocity $\langle \beta_t\rangle$ and $\langle u_t\rangle$ is given as 
\begin{equation}
\label{eq8}
\langle \beta_t\rangle=\frac{\langle u_t\rangle}{\sqrt{1+\langle u_t\rangle^2}} ,
\end{equation}

what calls for special attention is that although the $T$ arising from the non-extensive statistical theory can be different from usual temperature in Eq.(\ref{eq1}), the flow correction of the spectral temperatures is independent of the statistical model. In addition, the following functions will be used to study the collectivity of produced charmed meson in heavy-ion collisions~\cite{b5}:
\begin{equation}
\label{eq9}
\frac{d^{2}N}{2\pi m_Tdm_Tdy}=\frac{dN/dy}{2\pi T(m_0+T)} e^{-(m_T - m0)/T}, 
\end{equation}
where $m_0$ is the rest mass of the given hadron species. Such a method are used to analyze the charmed meson
spectra and to understand the radial flow collective velocity from the data in Pb-Pb collisions at 2.76 TeV. The function Eq.(\ref{eq9}) is applied to the spectrum of $\pi$, $K$, $p$($\overline{p}$), $\phi$, $\Lambda^0$, $\Xi$, $\Omega$ and $D^0$ for 0-10\% and 30-50\% centralities at 2.76 TeV. For $\Lambda^0$, $\Xi$ and $\Omega$, the semi-peripheral results are accomplished with 20-40\% centrality. The fitted parameters are shown in Fig.~\ref{fig3}.
The solid and hollow circles represent the results of the 0-10\% and 30-50\%, respectively. The slope parameters obtained by $D^0$ mesons and strangeness hadrons at the 0-10\% and 30-50\% are $0.113\pm 0.037$ and $0.107\pm 0.036$, respectively. And the linear fits of $\phi$, $\Lambda^0$, $\Xi$, $\Omega$ and $D^0$ data points in Fig.~\ref{fig3} show a smaller slope than $\pi$, $K$, $p$($\overline{p}$), indicating the former may freeze out earlier and obtain less transverse collectivity during the system evolution. All fits are performed up to $m_T - m_0 <$  1 GeV/$c^2$ for $\pi$, $K$, $p$($\overline{p}$), $<$ 2 GeV/$c^2$ for $\phi$, $\Lambda^0$, $\Xi$, $\Omega$, and $<$ 3 GeV/$c^2$ for $D^0$, respectively.

\begin{figure}[!htb]
\includegraphics
[width=0.9\hsize]{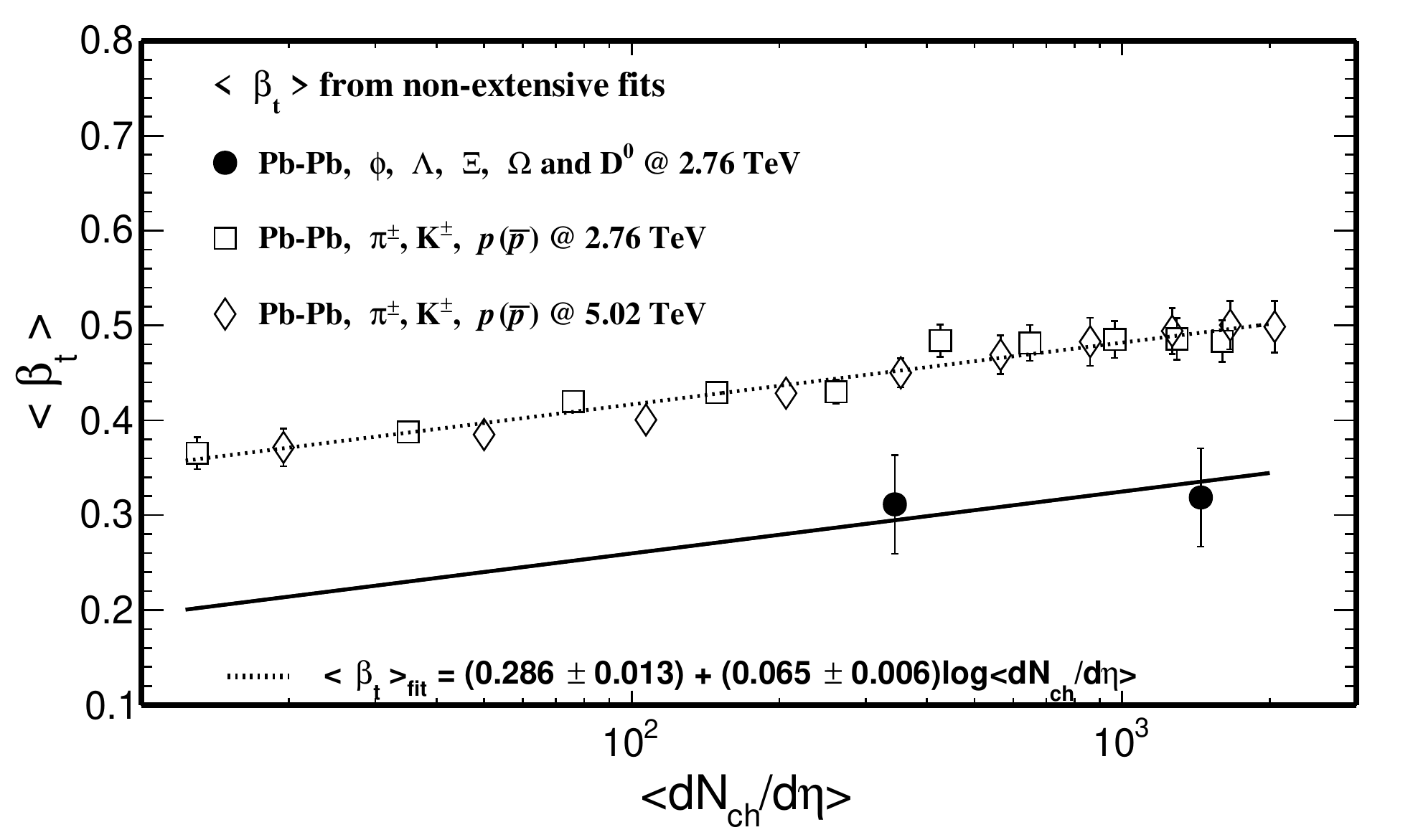}
\caption{The averaged radial flow velocity from Eq.(\ref{eq8}) as a function of the average event multiplicity. The solid circles represent $\phi$, $\Lambda^0$, $\Xi$, $\Omega$ and $D^0$ are combined to extract $\langle u_t\rangle$ at 0-10\% and 30-50\% centralities. The results of the light hadrons at 2.76 TeV and 5.02 TeV are shown by the hollow squares and hollow diamonds, respectively. The fit result is also shown at the bottom of the panel.}
\label{fig4}
\end{figure}

The fitted Tsallis-Pareto $T$ parameters in section~\ref{sec.3.1} are plotted as a function of the corresponding masses of $\pi^{\pm}$, $K^{\pm}$, $p$($\overline{p}$) at 0-5\%, 5-10\%, 10-20\%, 20-30\%, 30-40\%, 40-50\%, 50-60\%, 60-70\%, 70-80\%, and 80-90\% centrality bins in Pb-Pb collisions at 2.76 TeV and 5.02 TeV. The $\langle u_t\rangle$ values defined by Eq.(\ref{eq7}) are extracted, and the $\langle \beta_t\rangle$ distributions as a function of $\langle dN_{ch}/d\eta\rangle$ in Pb-Pb collisions at 2.76 TeV and at 5.02 TeV are shown by hollow squares and hollow diamonds in Fig.~\ref{fig4}, respectively. The $D^0$ is added to the process above at 0-10\%, 30-50\% centralities in Pb-Pb collisions at 2.76 TeV and is extrapolated to a lower centrality with the same centrality dependency as light flavor hadrons. We plot the relationship between $\langle \beta_t\rangle$ and $\langle dN_{ch}/d\eta\rangle$, where the result is shown by solid circles in Fig.~\ref{fig4}. We eventually obtain the thermal temperatures after the flow correction of the spectral temperatures for charmed mesons in section~\ref{sec.III}. 

Figure \ref{fig4} reveals that the multiplicity dependence of the averaged radial flow velocity is linear as a function of $\log \langle dN_{ch}/d\eta\rangle$. Furthermore, the averaged radial flow velocity increases with increasing multiplicity, which is consistent with the previous blast wave analysis~\cite{b17}. And the values can be extracted for $\pi^{\pm}$, $K^{\pm}$, $p$($\overline{p}$) by the non-extensive statistical theory:
\begin{equation}
\label{eq10}
\langle \beta_t\rangle=\left( 0.286\pm 0.013 \right) +\left( 0.065\pm 0.006 \right) \log \langle dN_{ch}/d\eta\rangle ,
\end{equation}
The linear dependence of the charmed meson is 
\begin{equation}
\label{eq11}
\langle \beta_t\rangle=\left( 0.129\pm 0.037 \right) +\left( 0.065\pm 0.000 \right) \log \langle dN_{ch}/d\eta\rangle ,
\end{equation}

\section{Results and discussion}\label{sec.III}

The measured transverse momentum distributions of charmed meson in pp($\overline{\text{p}}$) collisions from ALICE, CDF and STAR collaborations are fitted with Tsallis-Pareto distributions in Fig.~\ref{fig1}. The fit results are generally in good agreement with the data points. And the same Tsallis-Pareto fits are applied to transverse momentum distributions of charmed mesons in Pb-Pb collisions at 2.76 TeV and 5.02 TeV. Figure \ref{fig2} shows the measured charmed meson spectra together with the fit Tsallis-Pareto curves for different centrality bins in Pb-Pb collisions at 2.76 TeV (Fig.~\ref{fig2} (a)) and 5.02 TeV (Fig.~\ref{fig2} (b), (c), (d)).

\begin{figure}[!htb]
\includegraphics
[width=0.9\hsize]{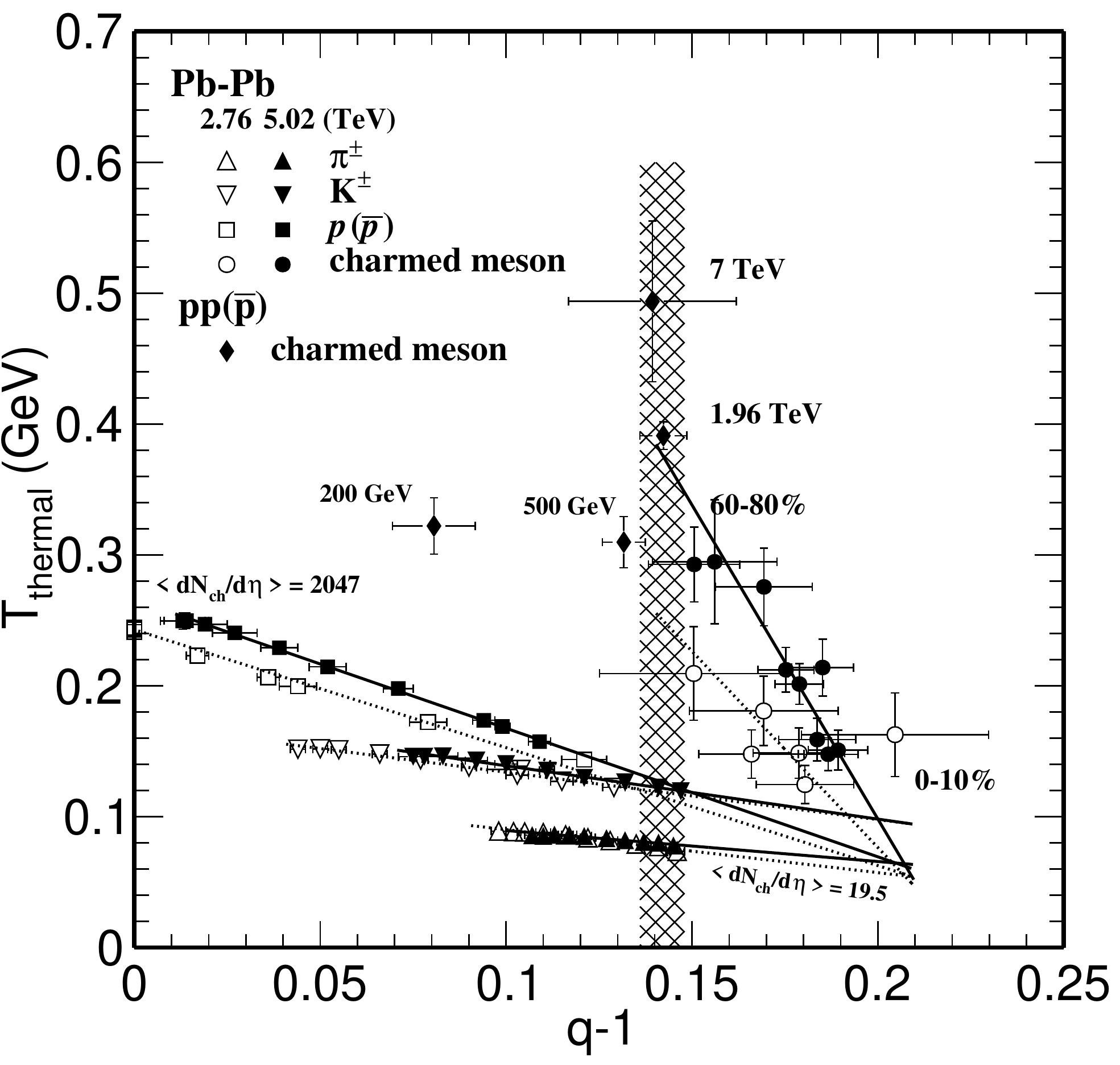}
\caption{Thermal temperature $T$ versus $q-1$, values of parameters from the Tsallis-Pareto fit to the identified particle spectra at different centralities in Pb-Pb (pp($\overline{\text{p}}$)) collisions at 2.76 TeV and 5.02 TeV (200 GeV, 500 GeV, 1.96 TeV, and 7 TeV), after transverse flow corrections. The vertical shaded-band marks the saturated value of $q-1$ = $0.142 \pm 0.010$ in pp($\overline{\text{p}}$) collisions with increasing energy. The solid and dotted lines are from Eq.(\ref{eq6}), the parameters are tabulated in TABLE~\ref{table2}.}
\label{fig5}
\end{figure}

The non-extensive feature $q \not=$ 1 has been observed from TABLE~\ref{table1}. To get a more physical understanding of this result, the dependence of the parameter $q$ on the size of the collisional system has been represented diagrammatically in Fig.~\ref{fig5}. In addition, the spectral temperature $T$, obtained from Tsallis-Pareto fit to the hadron spectrum, is larger than the original temperature $T_{thermal}$ by a blue shift factor due to the existense of a radial flow~\cite{b15}. The ordinate of Figure \ref{fig5} shows the thermal temperature values resulted from the flow correction of the spectral temperatures for different hadron species, and the flow correction formalism is given as:
\begin{equation}
\label{eq12}
T_{thermal}=T\sqrt{\frac{1-\beta_t}{1+\beta_t}}.
\end{equation}
%\vspace{-3mm}
whre $\beta_t$ is given by Eq.(\ref{eq10}) for $\pi ^{\pm}$, $K^{\pm}$, $p$($\overline{p}$) and Eq.(\ref{eq11}) for the charmed mesons. 

Figure \ref{fig5} shows the $T$ - ($q-1$) parameter space for $\pi ^{\pm}$, $K^{\pm}$, $p$($\overline{p}$), and charmed meson in Pb-Pb (pp($\overline{\text{p}}$)) collisions at 2.76 TeV and 5.02 TeV (200 GeV, 500 GeV, 1.96 TeV, and 7 TeV). To simplify this figure, the results for strange particles $\phi$, $\Lambda^0$ and $\Xi$ are not shown. The hollow and solid symbols represent Tsallis-Pareto parameter values at 2.76 TeV and 5.02 TeV, respectively. The $\langle dN_{ch}/d\eta\rangle$ = 2047 and $\langle dN_{ch}/d\eta\rangle$ = 19.5 represent the central and peripheral centrality bins for the light hadrons, respectively. There are five points around $T = 0.25$ at $q - 1 = 0$. This indicates that in such high energy the light flavor particles are totally thermalized in most central Pb-Pb collisions, thus their $q$ does not deviate from $1$ at high temperature. And the hollow and solid lines show the fitting results from Eq.(\ref{eq6}). The fits for charmed meson are performed in the range of 0.14 $< q-1 <$ 0.21 for 2.76 TeV and 5.02 TeV, and the fit parameters are tabulated in TABLE.~\ref{table2}. The charmed meson results for pp($\overline{\text{p}}$) collisions at different energy are displayed by diamonds. The vertical shaded-band marks the saturated value of $q-1$ = $0.142 \pm 0.010$ in pp($\overline{\text{p}}$) collisions with increasing energy. And the value is obtained by a constant fitting in the saturated region. To further investigate the $T$ - ($q-1$) correlations between pp($\overline{\text{p}}$) and Pb-Pb systems, we extend the fit lines along the q-1 reduction direction for charmed meson in Pb-Pb collisions. 

\begin{table}[]\footnotesize

\setlength{\tabcolsep}{-6mm}

\centering \caption{Values of parameters from linear fit Eq.(\ref{eq6}) to the $T$ - ($q-1$) correlations for $\pi ^{\pm}$, $K^{\pm}$, $p$($\overline{p}$), and charmed meson in Pb-Pb collisions. The uncertainties quoted are the errors returned from the fit.}\label{runResult}
\centering

\setlength{\tabcolsep}{.7mm}{

\begin{tabular}{ccccc}

\toprule
$\sqrt{s_{NN}}$ (GeV) & Hadron & E(GeV) & $\zeta ^2$ & $\chi ^2/ndf$ \\ \midrule

Pb-Pb, 2760         & charmed meson         & $3.000\pm 2.225$         & $0.226\pm 0.055$     & $2.11/4$         \\
         & $\pi^{\pm}$         & $0.329\pm 0.022$     &  $0.374\pm 0.017$      &  $7.27/8$        \\
         & $K^{\pm}$         &  $0.362\pm 0.014$     &  $0.470\pm0.016$        &  $15.91/8$        \\
         & $p$($\overline{p}$)         &  $0.901\pm 0.036$      &  $0.270\pm 0.011$   &   $11.04/8$       \\

Pb-Pb, 5020         & charmed meson         & $4.800\pm 1.460$         & $0.221\pm 0.015$     & $1.85/7$         \\
         & $\pi^{\pm}$         & $0.237\pm 0.027$     &  $0.478\pm 0.040$      &  $4.68/8$        \\
         & $K^{\pm}$         &  $0.407\pm 0.024$     &  $0.441\pm 0.018$        &  $2.53/8$        \\
         & $p$($\overline{p}$)         &  $0.982\pm 0.049$      &  $0.271\pm 0.010$          &   $0.62/8$       \\
\bottomrule

\end{tabular}

}
\label{table2}

\end{table}

\begin{figure}[!htb]
\includegraphics
  [width=0.9\hsize]{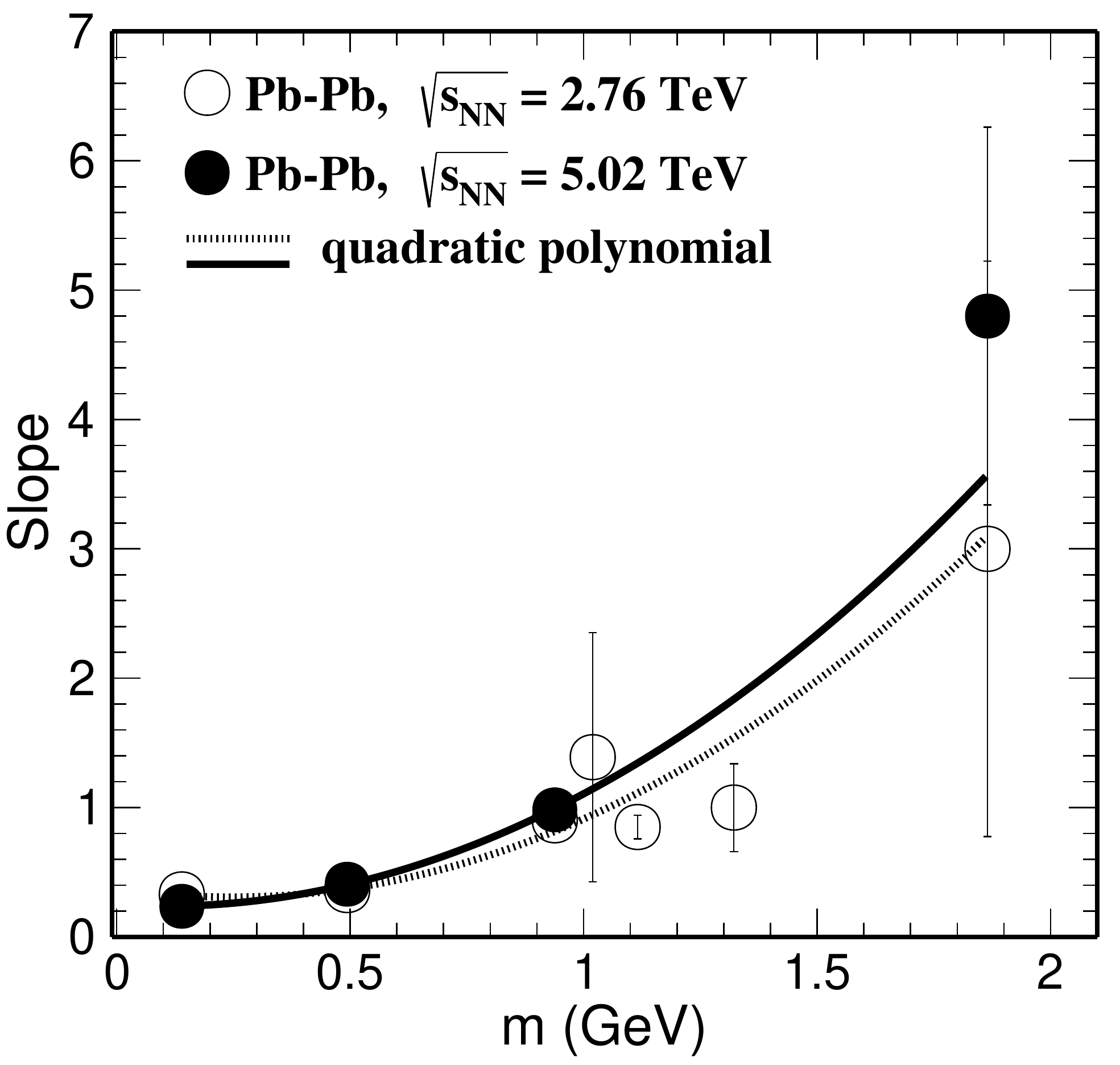}
\caption{The slope of the $T$ - ($q-1$) correlations as a function of hadron mass in Pb-Pb collisions at 2.76 TeV (hollow circles) and 5.02 TeV (solid circles). The curves are quadratic polynomial fitting.}
\label{fig6}
\end{figure}

We find that the Tsallis-Pareto distributions can provide satisfactory description for a wide range of transverse momentum dependence of charmed meson production in pp($\overline{\text{p}}$) and AA collisions over a wide energy scale. The value of the corresponding $\zeta ^2$ parameters in TABLE.~\ref{table2} can be used to test the strength of the correlation between $T$ and $q$. This correlation is inversely proportional to the mass of hadron on the whole. From TABLE.~\ref{table2} and Fig.~\ref{fig5}, several conclusions can be drawn as the followings.

(i) There is a significant linearity between thermal temperature and Tsallis-$q$ parameter for $\pi^{\pm}$, $K^{\pm}$, $p$($\overline{p}$) and charmed meson in Pb-Pb collisions at 2.76 TeV. And the slope of $T$ - ($q-1$) parameter is positively correlated with the hadron mass. The same conclusion can be obtained in Pb-Pb collisons at 5.02 TeV, and a clearer conclusion can be obtained due to larger statistics and more complete centrality bins.

(ii) The charmed mesons have a significantly higher slope than that of the light hadrons. The temperature of charmed mesons is found to be higher than that of light hadrons at the same $q-1$, indicating that heavy flavor requires a higher temperature to achieve the same degree of non-extensive as light flavors in the heavy-ion collision.  The slope of the same hadron is smaller at 2.76 TeV than at 5.02 TeV. In Fig.~\ref{fig6}, the slope of the $T$ - ($q-1$) correlations are plotted as a function of the hadron mass of $\pi^{\pm}$, $K^{\pm}$, $p$($\overline{p}$), $\phi$, $\Lambda^0$, $\Xi$ and charmed meson in Pb-Pb collisions at 2.76 TeV and 5.02 TeV shown as the hollow and solid circles, respectively. Data are fitted by quadratic polynomial function, and a deeper theoretical explanation is needed.

(iii) The charmed meson results for pp($\overline{\text{p}}$) collisions at different energy show that the thermal temperature increases with system energy while the $q$ parameter shows a saturated trend and stops at pp($\overline{\text{p}}$) limit, $q-1$ = $0.142 \pm 0.010$. And the results of most peripheral Pb-Pb collisions are found to approach the pp($\overline{\text{p}}$) limit, which suggests that less medium effect and more in similarity with pp($\overline{\text{p}}$) collisions are found in more peripheral heavy-ion collisions.

\section{Summary} \label{sec.IV}

In summary, we have presented fits for the transverse momentum spectra of $D^0$, $D^+$ and $D^{*+}$ mesons at mid-rapidity in Pb-Pb collisions at 2.76 TeV and 5.02 TeV. Similar analysis with non-extensive statistics has implemented to identified light hadron spectra for different centrality bins in Pb-Pb collisions at 2.76 TeV and 5.02 TeV after the flow corrections. Charmed meson production can be well described by Tsallis-Pareto distributions. In the $T$ - ($q-1$) parameter space, we observe that the slope has a positive dependence on hadron mass. And the temperature of charmed mesons is found to be higher than that of light hadrons at the same $q-1$, indicating that heavy flavor requires a higher temperature to achieve the same degree of non-extensive as light flavors in heavy-ion collisions. In addition, the slope distribution of $T$ - ($q-1$) correlations in Fig.~\ref{fig6} and the anti-correlation between thermal temperature and centrality for charmed meson need a deeper theoretical explanation.

For pp($\overline{\text{p}}$) collision system as a reference, we find that the thermal temperature increases with system energy while the $q$ parameter shows a saturated trend and stops at pp($\overline{\text{p}}$) limit, $q-1$ = $0.142 \pm 0.010$. Meanwhile, the results of most peripheral Pb-Pb collisions are found to approach the pp($\overline{\text{p}}$) limit, which suggests that less medium effect and more in similarity with pp($\overline{\text{p}}$) collisions are found in more peripheral heavy-ion collisions. And a uniform description for both small and large systems over a wide range of collision energy and hadron transverse momentum are found.

\end{document}